\documentclass[conference]{IEEEtran}
\usepackage{amsmath,amssymb,amsfonts}
\usepackage{algorithmic}
\usepackage{graphicx}
\usepackage{textcomp}
\usepackage{xcolor}

\usepackage{pifont}
\newcommand*\rot{\rotatebox{90}}
\newcommand*\OK{\ding{51}}

\ifCLASSOPTIONcompsoc
    \usepackage[caption=false, font=normalsize, labelfont=sf, textfont=sf]{subfig}
\else
\usepackage[caption=false, font=footnotesize]{subfig}
\fi
\usepackage[T1]{fontenc}

\usepackage[clock]{ifsym}

\PassOptionsToPackage{hyphens}{url}
\usepackage[hidelinks]{hyperref}
\usepackage[numbers]{natbib}
\newtheorem{question}{RQ}

\begin{document}

\title{Adapting Kubernetes controllers to the edge: on-demand control planes using Wasm and WASI}

\author{
\IEEEauthorblockN{Merlijn Sebrechts, Tim Ramlot, Sander Borny, Tom Goethals, Bruno Volckaert, Filip De Turck}
\IEEEauthorblockA{
IDLab, Department of Information Technology (intec), Ghent University - imec, Belgium
}
}

\IEEEoverridecommandlockouts
\IEEEpubid{\makebox[\columnwidth]{Pre-print of article accepted to IEEE CloudNet 2022] \copyright2022 IEEE \hfill}
\hspace{\columnsep}\makebox[\columnwidth]{ }}

\maketitle

\IEEEpubidadjcol

\begin{abstract}
Kubernetes' high resource requirements hamper its adoption in constrained environments such as the edge and fog. Its extensible control plane is a significant contributor to this, consisting of long-lived processes called "controllers" that constantly listen for state changes and use resources even when they are not needed. This paper presents a WebAssembly-based framework for running lightweight controllers on-demand, only when they are needed. This framework extends the WebAssembly System Interface (WASI), in order to run Kubernetes controllers as lightweight Wasm modules. The framework runs these Wasm controllers in a modified version of Wasmtime, the reference WebAssembly (Wasm) runtime, that swaps idle controllers to disk and activates them when needed. A thorough evaluation shows this framework achieves a 64\% memory reduction compared to traditional container-based controller frameworks.
\end{abstract}


\section{Introduction}
\label{chapter:intro}

The adoption of new technologies relies heavily on potential resulting cost savings. Financial and environmental cost reductions have been achieved by increasing computational density in cloud computing. Lower latency costs have stemmed from advancements in telecommunications, such as 5G, and the dispersion of cloud to fog, edge and IoT. These two innovation flows have independently led to widely accepted solutions. Combining these two ideas, however, poses a great opportunity. Adopting cloud computing techniques in the edge has the potential to greatly improve reliability, release software faster and expedite operations. Cloud computing utilizes cloud orchestration for optimal resource allocation and server cluster management such as Kubernetes~\cite{Brewer2015}, a Google-led open source project inspired by their internal Borg orchestrator~\cite{Verma2015, Burns2016}. Its extensible architecture parts a cluster in a control plane and a worker plane. Edge environments, which are typically large complex clusters with limited resources, can benefit significantly from having an orchestrator to manage complexity and size.

The primary orchestration targets of Kubernetes, however, are high-resource cloud clusters. Running Kubernetes on low-resource clusters suffers from relatively high control plane overhead costs, which hinders adaptation in the edge market segment. In complex cloud-native application deployments, operators \cite{Shubham2022} are used to automate actions on the Kubernetes cluster state, that would otherwise be performed by a human operator. These operators are one of the main cost drivers of the Kubernetes control plane. To react to changes in the Kubernetes cluster state, the operators have to run as long-living processes. Even if the operator’s control loop is idle, the container and process still use cluster resources. For complex applications that use many operators, these overhead costs quickly accumulate and account for a significant portion of the resource utilization. This is especially problematic for low-resource deployments.

Since global edge application configuration and deployment is a complex task, it is often abstracted by the service provider and included in a Function as a Service (FaaS) offering.
FaaS applications are better suited for low-resource edge environments thanks to their fast on-demand scaling properties. Specifically, some edge FaaS platforms use WebAssembly (WASM), a browser technology designed as a portable binary code format that can be assembled from a range of programming languages and that is well suited to resource-constrained environments.
On edge FaaS platforms, like Cloudflare Workers \cite{Cloudflare2022} and Fastly Compute@Edge \cite{Fastly2022}, WebAssembly is used to securely isolate workloads with reduced overhead and scale-to-zero capabilities.

This research aims to make Kubernetes more suitable to resource-constrained edge environments by running its control plane in a FaaS platform based on WebAssembly. Specifically, the research aims to answer the following questions:

\question{How can Kubernetes use WebAssembly to run parts of its control plane?}

\question{How does running the Kubernetes control plane in WebAssembly impact its overhead?}

\question{What situations affect the overhead difference between WebAssembly and regular operators?}

\question{What is the effect on resource usage of running this control plane in an on-demand manner on a FaaS platform?}

Section~\ref{chapter:literature} investigates the state of the art concerning adapting Kubernetes to the edge. Section~\ref{chapter:architecture} explains the architecture of our WebAssembly-based operator solution and Section~\ref{chapter:implementation} discusses how this architecture is implemented. In Section~\ref{chapter:result}, our benchmark results of the WebAssembly runtime are discussed, as well as the methodology to achieve these results.

\section{Related Work}
\label{chapter:literature}

WebAssembly is used by a number of large edge computing platforms such as Cloudflare Workers \cite{Cloudflare2022} and Fastly Compute@Edge \cite{Fastly2022} to securely isolate workloads with reduced overhead and scale-to-zero capabilities. Although WebAssembly runtimes generally incur a performance hit compared to native code, this is not inherent to WebAssembly itself, as some runtimes such as Wasmachine show performance \textit{improvements} of up to 21\% compared to native code~\cite{wen_wasmachine_2020}. WebAssembly is especially useful in a Serverless or FaaS setting because of its quick startup times. The Sledge serverless platform, for example, uses WebAssembly to achieve up to 4 times higher throughput and 4 times lower latencies compared to the state of the art~\cite{gadepalli_sledge_2020}.

As more and more computation moves towards the edge, so do supporting components such as Kubernetes~\cite{carvalho_edge_2021}. Efforts to reduce the footprint of Kubernetes in the edge have shown some success~\cite{kjorveziroski_kubernetes_2022}. Although it is possible to reduce the footprint of the existing Kubernetes codebase, the best performance improvements require a complete re-implementation of core Kubernetes components~\cite{goethals_fledge_2020}. The overhead of Kubernetes is larger than its core implementation, however, as many microservice deployments and frameworks use custom controllers for management, which add significant memory strain.

Recent additions to the Kubernetes ecosystem, like Krustlet~\cite{microsoft_research_introducing_2020} and Wasmedge~\cite{wasmedge_developers_wasmedge_2022}, add support for WebAssembly workloads as an alternative to OCI containers. Efforts to use WebAssembly for extending control-plane components is also underway, such as in the Kubewarden project, which allows writing custom Kubernetes policies in any language and compile them to WebAssembly.~\cite{kubewarden_project_authors_kubewarden_2022} Finally, some strides have been made towards using WebAssembly for writing \textit{any} control logic, by creating Wasm Operators~\cite{guardiani_kubernetes_2020}. This effort, however, still requires controllers to continuously run, regardless of whether any changes need to be processed. Thus, a solution is needed which combines Kubernetes controllers with WebAssembly and Serverless in order to have a truly event-based Kubernetes control plane which only activates when needed and gets unloaded after execution.

\section{Solution architecture: WASM operator}
\label{chapter:architecture}

Because of Kubernetes' can-always-fail design, an operator application is not supposed to hold any internal state across reconciliation iterations except for caches. The operator generally uses the Kubernetes API to store state. This theoretically allows running each iteration of the reconciliation loop without storing state in between. This property also holds for FaaS systems, where there is no guarantee for state preservation in between function calls. FaaS solutions for constrained edge environments often utilize Software-based Fault Isolation (SFI) instead of process isolation. WebAssembly lets you create SFI applications based on code written in existing high-level languages.

%
\begin{figure}[h]
    \centering
    \includegraphics[width=0.85\linewidth]{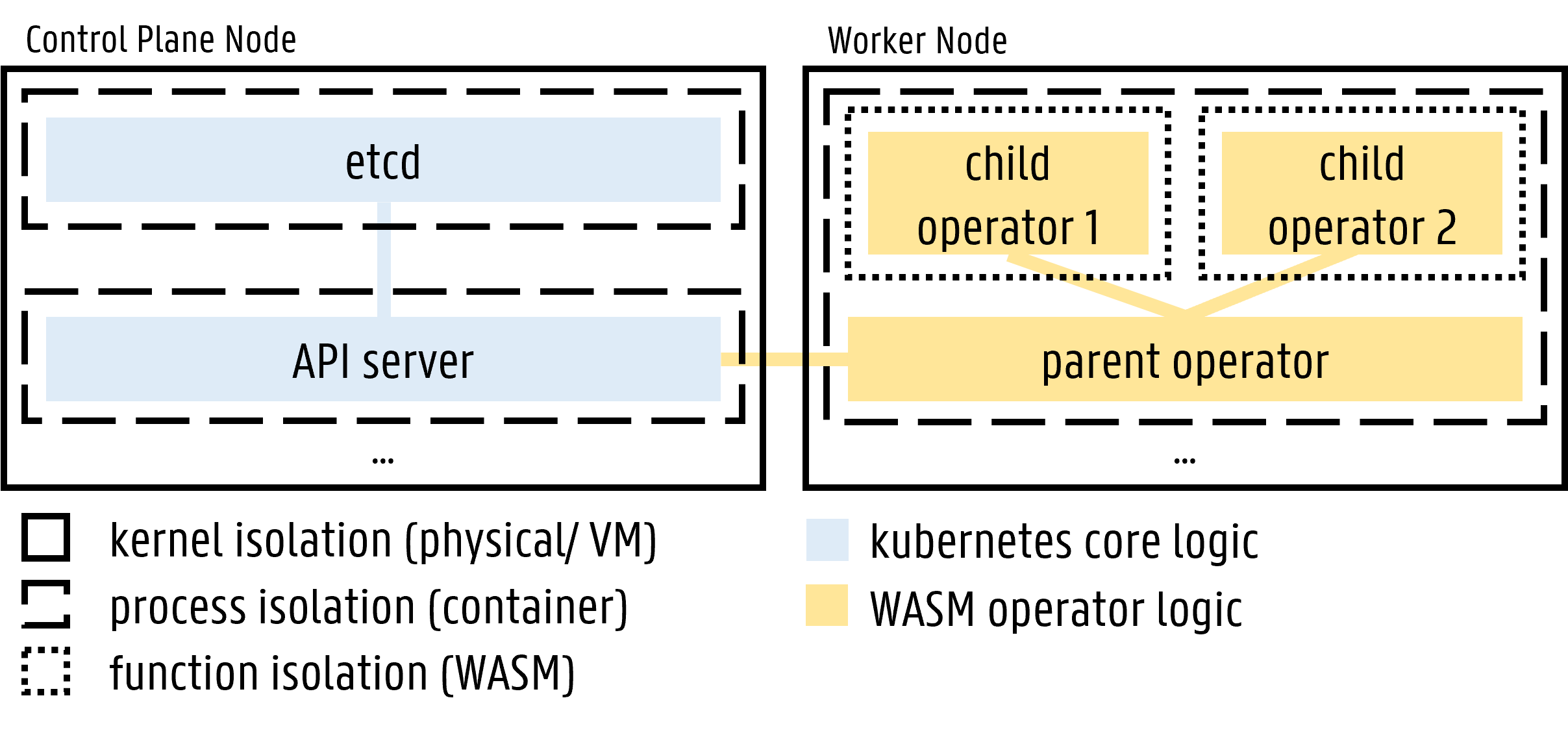}
	\caption{Design of our WASM operator architecture.}
    \label{fig:wasm_operator_design}
\end{figure}
The \textit{WASM operator architecture} presented in this paper and visualized in Figure~\ref{fig:wasm_operator_design} attempts to realize all the beneficial aspects that solutions in prior work have achieved \cite{Srinivas2022, Yeh2022, Ferris2019}.
The main parts of the architecture are the parent operator and the child operators.
The child operators run as WASM instances in the WASM runtime embedded in the parent operator.
We use an existing WASM runtime implementation as embedded runtime.
The beneficial aspects are listed below.
\begin{itemize}
    \item \textbf{Isolation overhead:} All child operators run in the same process as the parent operator. This process runs inside a single container in a single Kubernetes pod.
    Isolation is provided by the WASM engine, eliminating the overhead due to container isolation.
    \item \textbf{Modularity: } The WASM runtime makes it possible to add or remove child operators without interfering with the other active child operators.
    \item \textbf{Simple child operator:}
    In our architecture, the parent operator extends the WASM runtime with host functions that can be used by the child operators to communicate with the Kubernetes API. Low-level operator logic is moved to the parent operator. This reduces the complexity and overhead of the client operators.
    \item \textbf{Scale-to-zero:} To limit the overhead of inactive operators, our architecture allows to dynamically unload inactive operators.
\end{itemize}

In order to efficiently make Kubernetes API requests, we want child operators to perform them asynchronously.
Existing WASM runtimes offer no support for asynchronous calls or offer a solution that is incompatible with idle module unloading. Therefore, we created a new solution that adds support for asynchronous operations to the WASM runtime that is embedded in our parent operator.
By extending the WASM runtime, we allow the child operators to wait for host functions asynchronously.

\subsection{Parent operator asynchronous runtime}
\label{architecture:parent_operator}
\begin{figure}[h]
    \centering
    \includegraphics[width=1\linewidth]{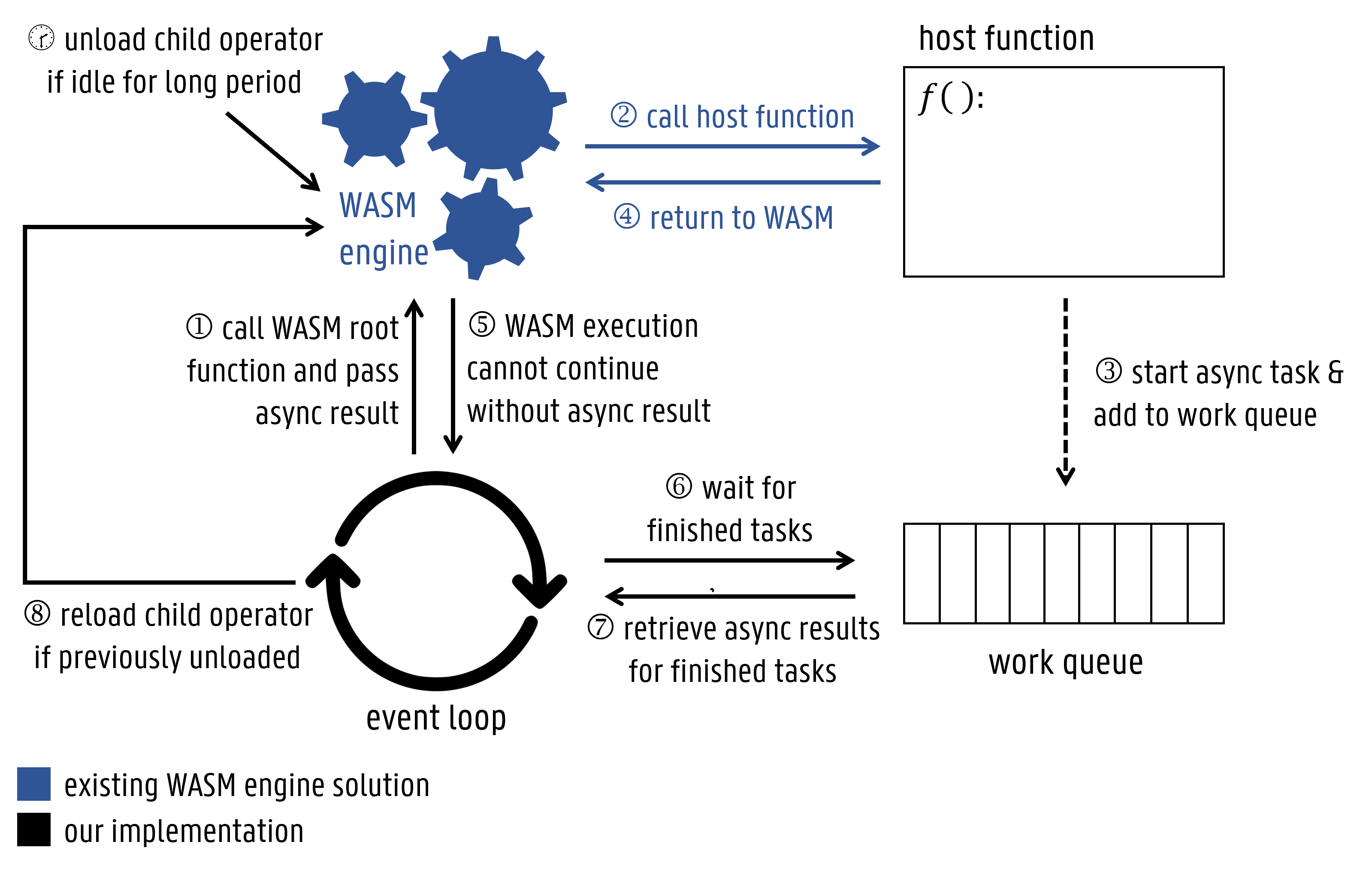}
	\caption{The design of the parent operator incorporates a WASM runtime event loop which repeatedly performs actions 1-8; making it possible to asynchronously call host functions from within a WASM instance.}
    \label{fig:wasm_runtime_event_loop}
\end{figure}
Figure~\ref{fig:wasm_runtime_event_loop} shows how the parent operator manages the asynchronous operations of a child operator and unloads an inactive child operator after a long period of inactivity. The main components of the parent operator are the WASM engine, the host functions exposed to the WASM instance and the work queue. For the WASM engine component and some of the host functions, existing solutions can be used.

The solution works as follows: Each WASM module has an entry point that executes the main function in the child operator, shown in Figure~\ref{fig:wasm_runtime_event_loop} as {\large \textcircled{\small 1}}. Environment interactions happen through the calling of WASM host functions {\large \textcircled{\small 2}}. Some of these actions are asynchronous and do not directly yield a result. These asynchronous actions are started and added to the work queue {\large \textcircled{\small 3}}, directly returning control to the WASM module {\large \textcircled{\small 4}}. After executing all synchronous logic, the WASM execution stops and the control is returned to the event loop {\large \textcircled{\small 5}}. This loop checks if any of the actions in the work queue finished {\large \textcircled{\small 6}} and passes the results of that finished action {\large \textcircled{\small 7}} back to the WASM engine {\large \textcircled{\small 1}}, reloading the child operator in case it had been previously unloaded {\large \textcircled{\small 8}}. When returning to WASM, a new set of synchronous actions are performed by the engine. Long-running operators repeat this process indefinitely. These operators are always waiting for new asynchronous inputs, like events in a watch stream.

The runtime might detect that a certain child operator has not been receiving any asynchronous results over a long period (marked as \showclock{2}{30}). This is indicative for an operator that reached a steady state in its reconciliation process. Most likely, it will only restart its logic after external applications changed the state of the Kubernetes resources that it manages. This could mean that the operator remains idle for multiple hours. In such cases, it can be more resource-efficient to unload and swap the WASM instance to disk.

\subsection{Child operator asynchronous client}
\label{architecture:child_operator}

All client operators run as single-threaded asynchronous WASM instances. The child operator is started by the host which calls the \texttt{start} function that is exposed by the WASM module. This initial function starts the operator reconciliation loop, which makes asynchronous requests. These asynchronous futures \cite{Cramer2022} are \texttt{awaited} by the child operator, but some of these futures \texttt{await} asynchronous host function results from the parent runtime. If the child operator cannot continue without new results from the host environment, it stops the execution and returns to the host. If none of the pending asynchronous requests have finished already, the host waits for one of them to finish, as described in Section~\ref{architecture:parent_operator}. Once a request finishes, the host returns the result to the child operator, such that the child operator can finish the linked asynchronous request. This restarts the whole process.

\section{Implementation}
\label{chapter:implementation}

The latest version of our implementation and the scripts used for the end-to-end tests can be found on Github~\cite{Ramlot2022}.

\subsection{Prior work}

Our operator implementation builds upon the proof of concept (PoC) made by Francesco Guardiani and Markus Thömmes~\cite{Guardiani2020}. This PoC provides a WASM operator solution based on the Wasmer~\cite{Akbary2022} WASM runtime and a hacked version of the kube-rs~\cite{Albrigtsen2022} library. However, at the time of writing, it has been 2 years since this project was updated. Since the API of the Wasmer runtime drastically changed after its v1 release, and the hacks applied to the kube-rs project are not well documented, upgrading the PoC was not straight forward. Furthermore, the Wasmer project lacks the future potential that other open-source initiatives, like Wasmtime, can offer. To update kube-rs more easily in the future, a new project structure was required. Moreover, the original version of the PoC cannot unload inactive operators as its architecture is different from the architecture proposed in Section~\ref{chapter:architecture}. We refactored the PoC and updated it to implement the aforementioned architecture. Finally, we implemented several improvements to further optimize the PoC implementation, such as adding support for caching compiled WASM modules for later reuse.

\subsection{Parent operator: WASM runtime}
\label{impl:parent_operator}

The parent operator extends the Wasmtime WASM runtime. Wasmtime was chosen over other WASM runtimes, because it is the flagship WASM engine from the Bytecode Alliance, with support from some of the biggest players in the technology industry. Our implementation configures Wasmtime to compile ahead of time (AOT) new WASM modules to machine code to eliminate the compiler memory overhead at runtime. These compiled modules are cached on disk and can be reused when possible. To initiate these compiled modules, Wasmtime only has to map the file to memory and provide the necessary tools to communicate with this initiated module. Because of the use of file-backed memory, for idle operators, these memory locations can be dropped from memory by the kernel when needed. If the memory region needs to be accessed again, a page-fault will be triggered, and the kernel will load the file back into memory. However, the dynamically populated memory of the WASM module will not be unloaded automatically from memory. That is why our implementation adds a custom unloading and disk swapping implementation in the parent operator. This makes unloading and swapping possible, even on systems without swap enabled at operating system level.

\subsection{Parent operator: host functions}
\label{impl:host_functions}

\begin{figure}[ht]
    \centering
    \includegraphics[width=0.70\linewidth]{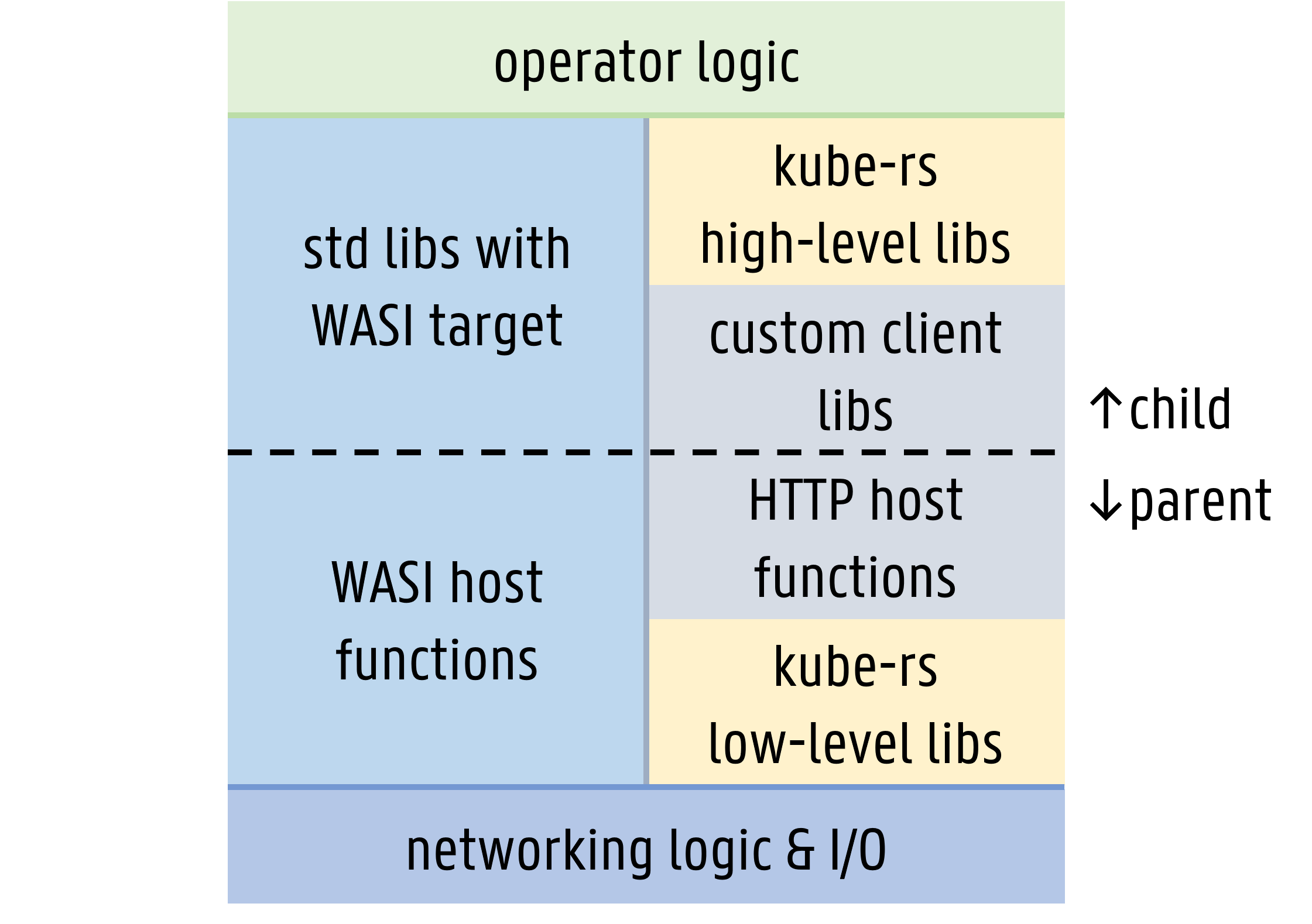}
	\caption{The operator libraries are split between the parent and child operator and consist out of existing WASI libraries and our own custom libraries based on kube-rs.}
    \label{fig:implementation}
\end{figure}

WASM host functions are functions exposed by the WebAssembly runtime to the WASM instances. The Web Assembly System Interface (WASI) is a standardized set of these host functions. Our implementation can benefit from the existing Wasmtime library that readily implements these WASI host functions, reducing the implementation and maintenance burden of our solution. A core aspect of Kubernetes operators is communicating with the Kubernetes API server. However, at the time of writing, the WASI spec has not yet standardized sockets as part of the interface \cite{Bakker2022}. This means that for our implementation, we had to implement custom HTTP host functions to create a working WASM operator setup. As shown in Figure~\ref{fig:implementation}, our implementation uses the low-level part of kube-rs for the Kubernetes host function implementation.
The high-level kube-rs functionality is implemented in the child operator. The added host functions are asynchronous functions, meaning that they return control to the WASM module immediately, while returning an \texttt{async\_id} that references a task in the work-queue as described in Section~\ref{architecture:parent_operator}.

\subsection{Child operator: client libraries}
\label{impl:child_operator}

All Kubernetes operator domain knowledge is implemented in the custom reconciliation loops, that are defined in the child operators. Our language preference for the child operators is Rust since the Rust standard libraries best support the WASI host function calls. Golang, which is normally used in Kubernetes, has no support for WASI in its default compiler. Additionally, Golang is a garbage collected language, which have been shown to use more memory \cite{Hertz2005}. Another advantage of choosing Rust as language is that an easier interoperability between the parent and child operator is obtained. The (de)serialization logic mentioned in Section~\ref{impl:host_functions}, can be reused for both the parent and child, since they are both implemented in Rust.

\section{Resource utilization}
\label{chapter:result}

\subsection{Test setup}


\begin{figure}[h]
    \centering
	\includegraphics[width=0.9\linewidth]{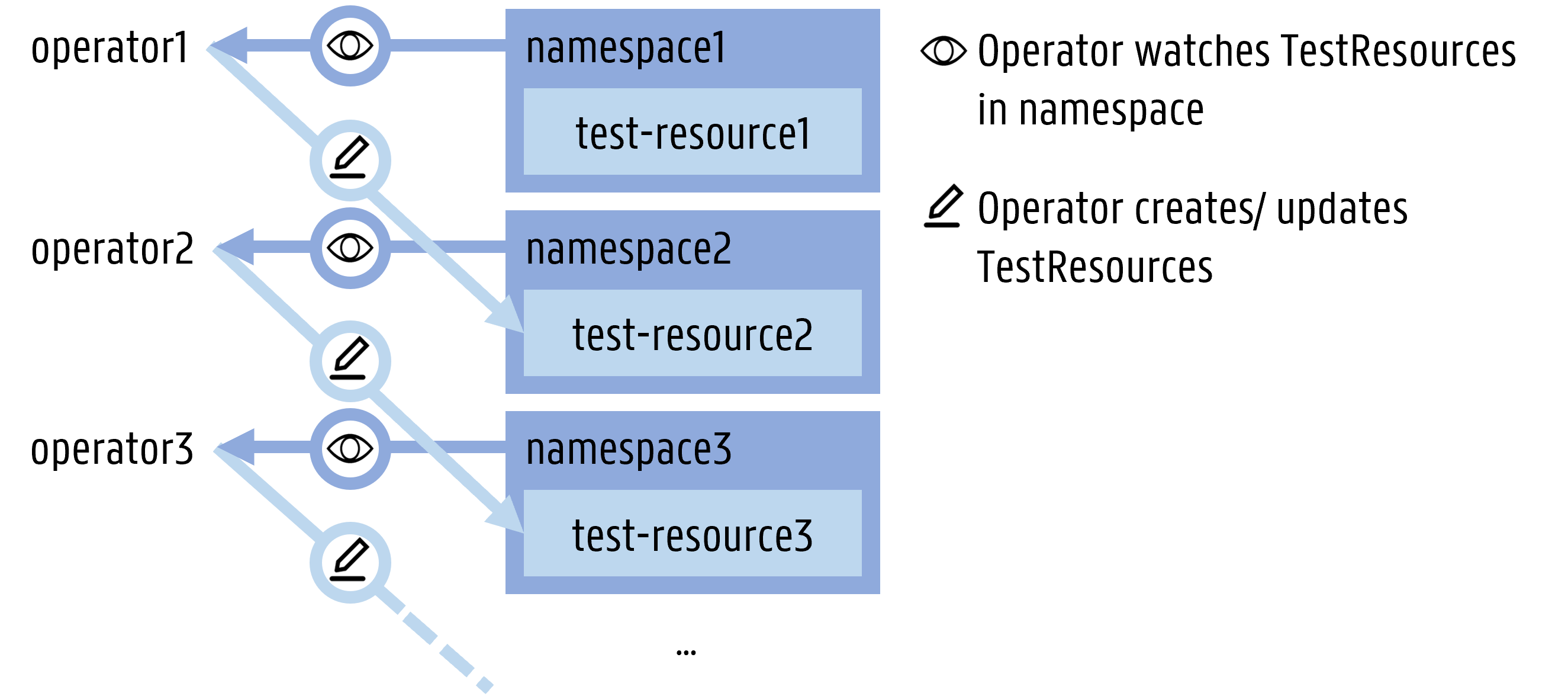}
	\caption{The test setup for the synthetic-operator workload, each operator is responsible for the propagation of changes from one namespace to another.}
	\label{fig:synthetic_operator_test_setup}
\end{figure}

The test synthetic-operator, as shown in Figure~\ref{fig:synthetic_operator_test_setup}, simulates a workload with $N$ different operators, which depend on each other's actions and are idle for most of the time. Each operator watches a namespace for \texttt{TestResources} and only reconciles, once a resource is created or updated. It then updates/ creates the resource in its destination namespace. For a full update of all resources, all operators must update their resource one-by-one. This means the full end-to-end latency equals the accumulated individual operator latencies. This synthetic workload simulates a highly dependent and interactive operator setup.

\label{result:memory_usage_method}

Measuring the memory footprint of a workload execution, requires accounting all the memory usage effects that the process has on the system. This is a non-trivial problem.
The memory utilization measure that we use is determined by limiting the memory usage, as determined by cgroup v2 \cite{Heo2015}, until the application is being slowed down as determined by the Linux PSI metric \cite{Weiner2018}.
For each run, we determine a upper bound memory limit.
Each upper bound is defined as the memory limit that is not exceeded for 95\% of the selected time range duration.
For each configuration, which is defined by an operator type and number of operators, five independent runs were performed, each yielding one upper bound for the active and one for the idle period.
Per operator type, we tested the number of operators from 10 to 100, in increments of 10.
For the active and idle selection separate, based on the resulting 50 upper bounds for each operator type, we trained a linear regression model.
Using this linear model, we determined the 95\% prediction interval in which we expect with 95\% certainty the upper bound memory usage of a new run with the given configuration, as described by Neter et al. \cite{Neter2005}.

\label{result:e2e_latency_method}

The end-to-end latency is measured by the synthetic-operator test for the active period of the test. Each set of reconciliations starts from an update of the \texttt{TestResource} in namespace $1$ until the \texttt{TestResource} in namespace $N$ is updated. The time from start to end is measured and each reconciliation set is repeated 500 times per run, resulting in 500$N$ reconciliation iterations.
As described in Section~\ref{result:memory_usage_method}, for each configuration, which is defined by an operator type and a number of operators, five independent runs are performed.

\subsection{Golang container, Rust container and Rust WASM compared}
\label{result:languages_memory_usage}

\begin{figure}
    \centering
    \subfloat[active]{
        \includegraphics[width=0.47\linewidth]{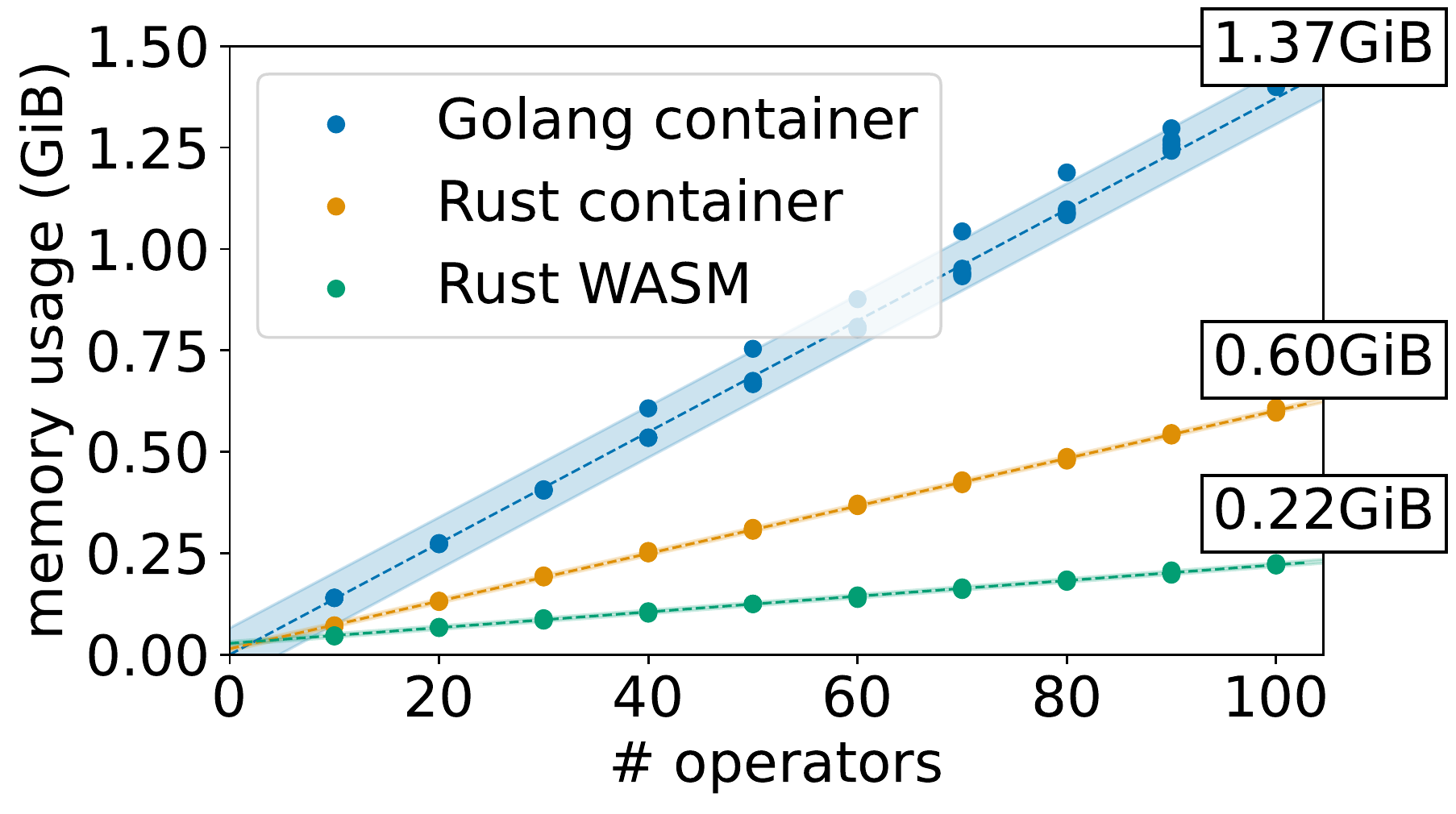}
        \label{gra:languages_active}
    }
    \subfloat[idle]{
        \includegraphics[width=0.47\linewidth]{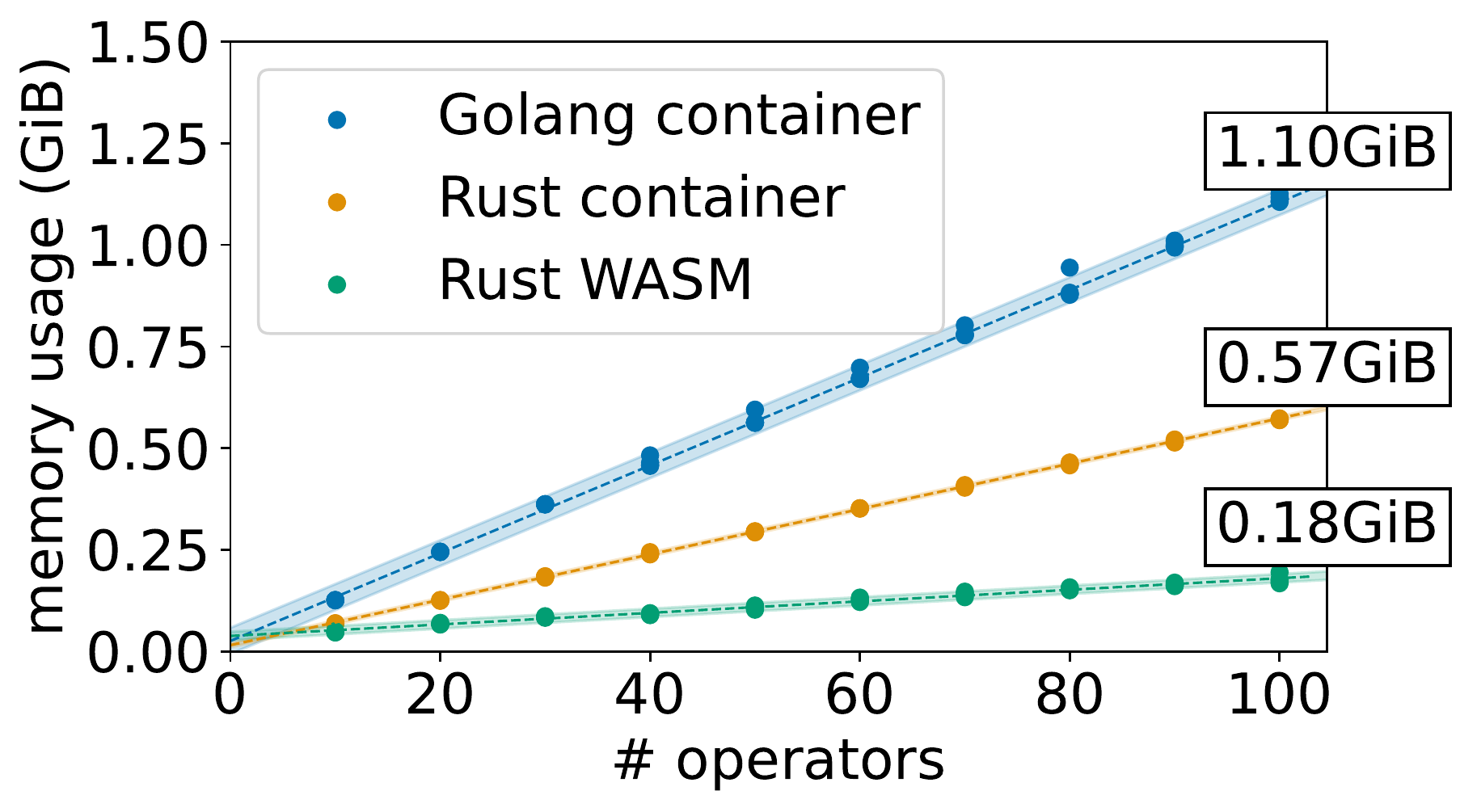}
        \label{gra:languages_idle}
    }
    
	\caption{
	    The memory 95\% upper bounds of the different languages/ isolation techniques are ordered as follows: Rust WASM < Rust container < Golang container; all operators use less memory when idle.
    }
    \label{gra:languages}
\end{figure}

Figure~\ref{gra:languages} shows the obtained memory upper bounds for container-isolated operators written in Golang and Rust and a WASM-isolated Rust operator. The colored areas represent the 95\% prediction intervals for the regression models as described in Section~\ref{result:memory_usage_method}.

Figure~\ref{gra:languages_active} shows the results for the active period. The Golang-based operator clearly uses the most memory. For 100 active operators, switching from Golang to Rust resulted in a $56.06\%$ upper bound memory reduction. WASM operators even yielded an $83.81\%$ reduction compared to Golang operators. Compared to Golang, the Rust operators use entirely different operator library and framework implementations. Each implementation has its own memory trade-offs, which can lead to large differences in memory usage. Additionally, as discussed in Section~\ref{result:memory_usage_method}, garbage collected languages like Golang, typically are less memory efficient than languages without garbage collector like Rust. The Rust container-based operator and the WASM-based operator share much of their source code. However, the WASM-based operators use less memory than container-based operators. This is due to the reduced complexity of the WASM child operator, as much of its low-level operator logic is moved to the parent operator. Moreover, the different isolation techniques used result in a net reduced isolation overhead, which is further explored in Section~\ref{result:cost_of_isolation}.

Figure~\ref{gra:languages_idle} shows that, as expected, all operator types utilize less memory in case of idle workloads, we observed a $14.21\%$ reduction on average.
Compared to 100 idle Golang operators, 100 idle Rust operators utilized $48.04\%$ less memory, which is a smaller reduction than when comparing active operators. However, 100 idle WASM operators still used $83.65\%$ less memory compared to idle Golang operators, similar to the active situation. The smaller reduction in memory usage of container-based Rust operators versus Golang operators is due to Golang experiencing a higher relative reduction in memory consumption when going from active to idle. Based on the typical usage pattern of an operator, which can be idle for a long period of time, it is clear that idle memory usage is important.


\begin{figure}
    \centering
	\includegraphics[width=0.85\linewidth]{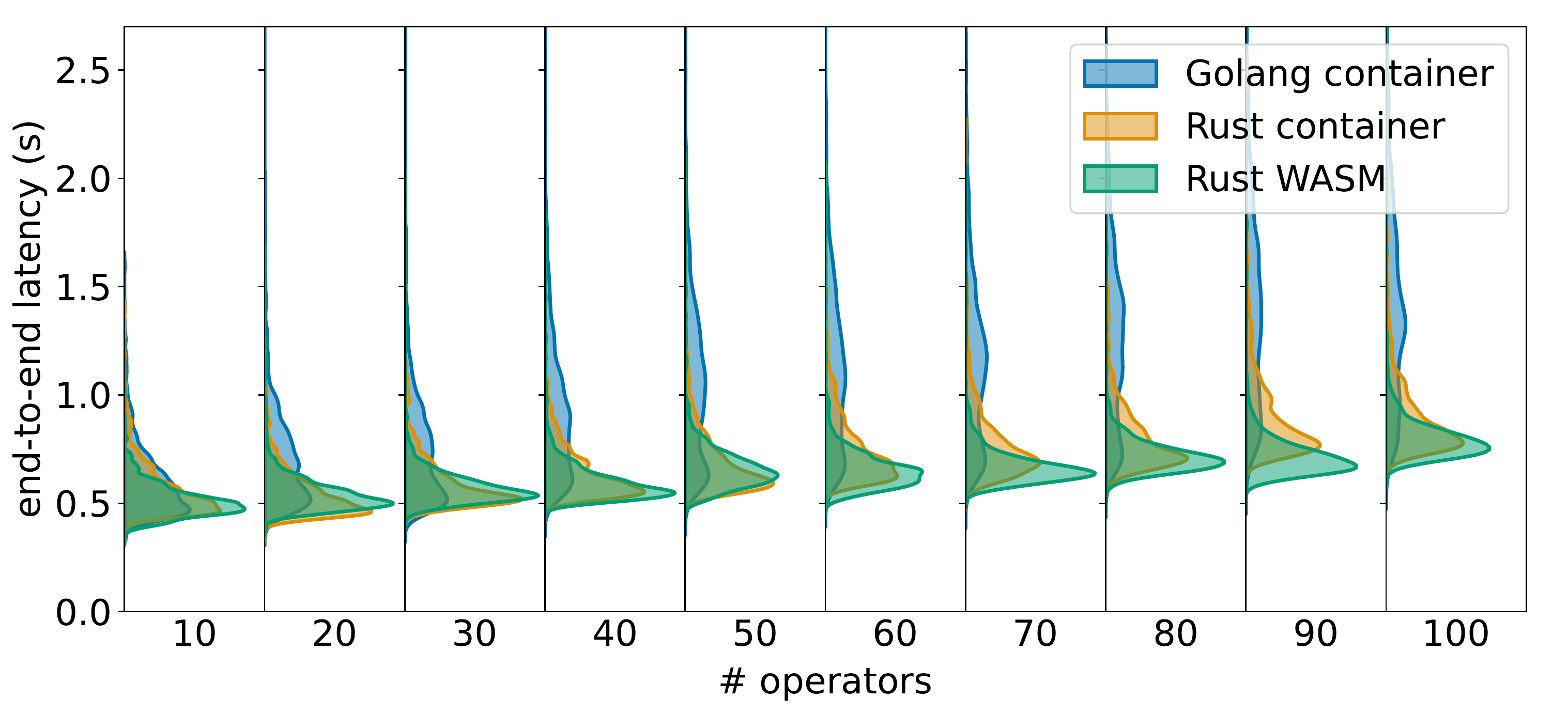}
	\caption{The end-to-end latency of WASM operators is identical to Rust operators.}
	\label{gra:languages_e2e}
\end{figure}

In Figure~\ref{gra:languages_e2e}, the obtained latency distributions for the different operator types are displayed, which were obtained as described in Section~\ref{result:e2e_latency_method}. Based on Jangda et al. \cite{Jangda2019}, WASM performance can be 2.5x slower worst-case compared to native execution. The WASM version of the synthetic-operator, however, did not experience any latency penalties. The latency for the Golang operator increased more than the other operators with increased number of operators. However, this is most likely due to the memory pressuring algorithm that adds more latency to Golang because its less memory efficient. There was no measured useful difference in the average latency between the WASM and Rust implementations that was greater than the measured noise. The main bottleneck in the operator's execution is I/O. Therefore, the latencies that occur in CPU-heavy workloads do not affect the synthetic-operator workload much.

\subsection{Cost of isolation}
\label{result:cost_of_isolation}

\begin{figure}
    \centering
    \subfloat[active]{
        \includegraphics[width=0.47\linewidth]{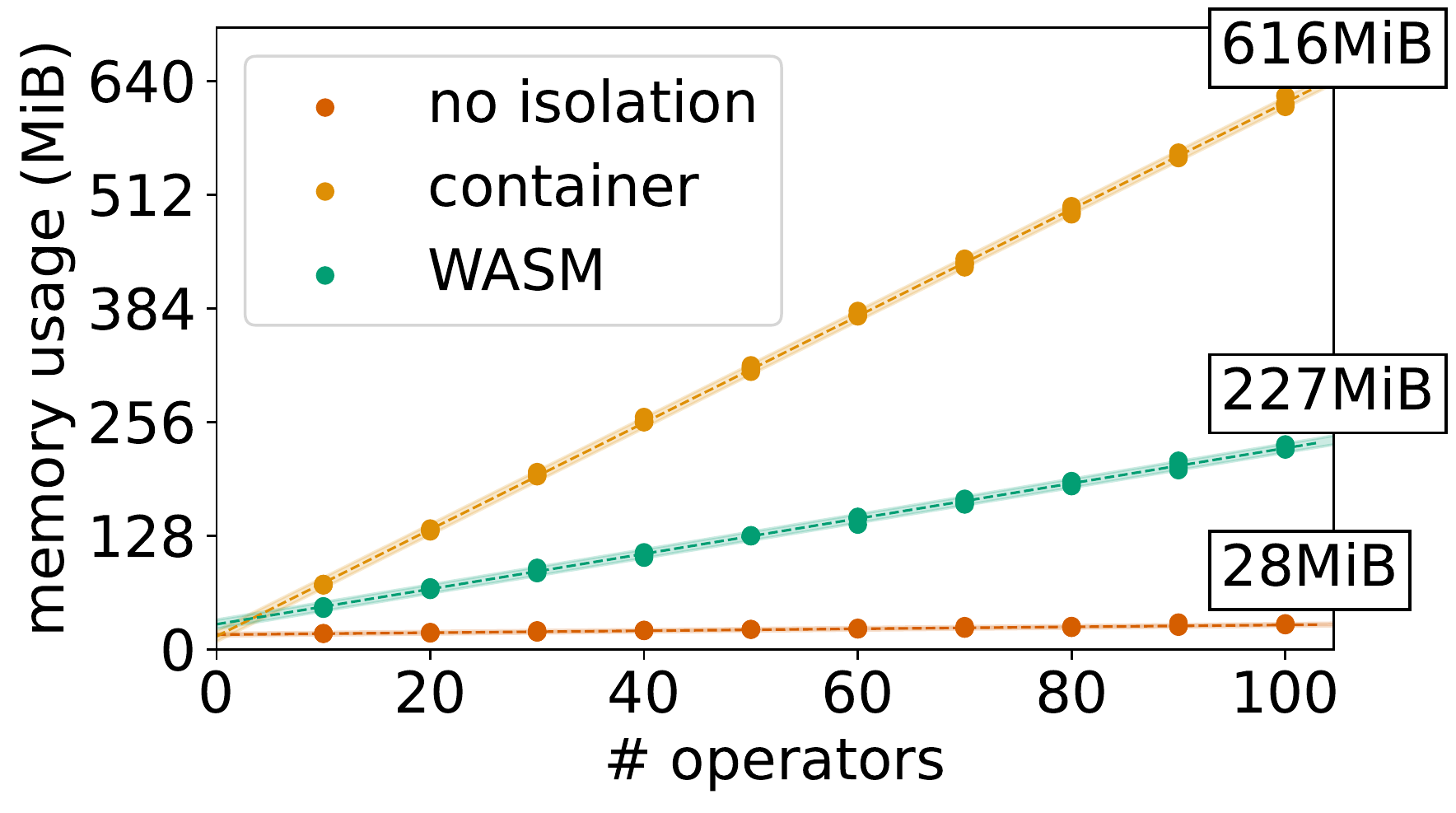}
        \label{gra:dynamic_active}
    }
    \subfloat[idle]{
        \includegraphics[width=0.47\linewidth]{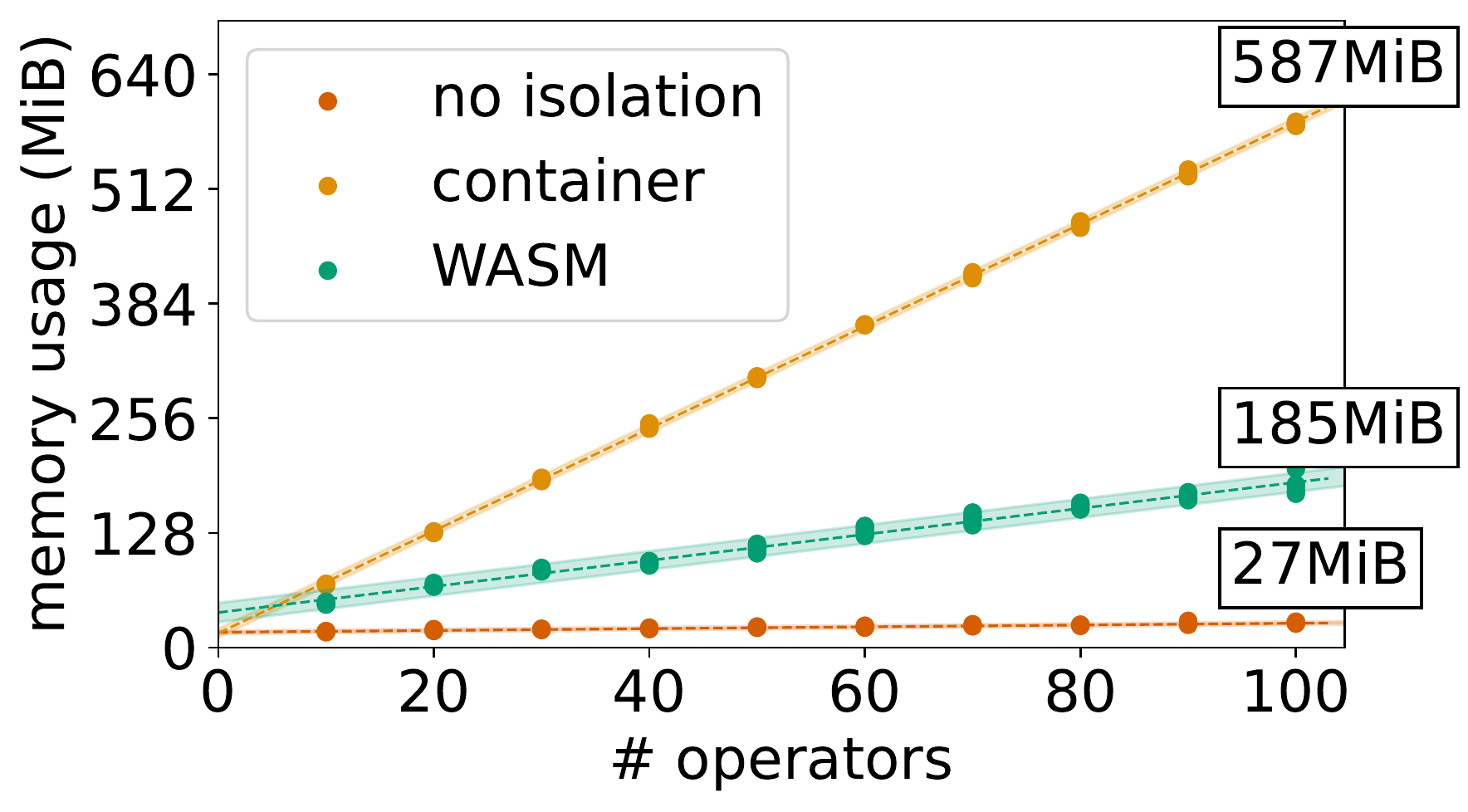}
        \label{gra:dynamic_idle}
    }
    \caption{The memory 95\% upper bounds of non-modular, container-modular and WASM-modular operators show that WASM outperforms container-based isolation, but additional improvements are possible since having no isolation is still much more efficient.}
    \label{gra:dynamic}
\end{figure}

Figure~\ref{gra:dynamic} shows the obtained memory upper bounds for Rust operators using no isolation, using containers and using WASM. The colored areas represent the 95\% prediction intervals for the regression models as described in Section~\ref{result:memory_usage_method}.

The solution with no isolation is the most resource efficient. This operator is able to scale to 100 control loops without significant additional memory overhead. Both the WASM-based and container-based setups experience significant per-operator overhead.
Additionally, the WASM-based operator has a higher initial constant memory overhead.
However, since the container-based solution performs worse per-container, this initial overhead can be compensated. In case of the active situation, the WASM-based solution is more memory efficient than the container-based solution with 95\% certainty starting from six operators. For the idle operators, this starts from eight operators.

The container-based operators are managed by Kubernetes and each run in a separate Kubernetes pod. Our Kubernetes setup uses containerd \cite{Crosby2022} to manage the containers. In our tests, the biggest overhead contributor was the per-pod \textit{containerd-shim} process which equates to about 5MiB per pod.
The WASM runtime can isolate the modules without introducing such a big overhead. Instead, it introduces a constant initial overhead that does not depend on the number of operators.
This memory overhead is due to the WASM runtime, including the low-level operator logic.

Our tests showed that a major memory usage reduction can be achieved by using no isolation.
However, having no isolation between operators means that all operators should be fully trusted even for not having errors. Additionally, it results in a lack of modularity: it is not possible to dynamically add or remove controllers. In an operator design based on Kubernetes pods, operators can be added and removed dynamically. Also, WASM modules can be loaded dynamically by the parent operator, without having to restart the parent operator process. WASM is a good intermediate solution, providing isolation and modularity while still being more memory efficient than the container-based solution.

\subsection{Automatically unloading WASM modules}
\label{result:unloading}
\begin{figure}
    \centering
    \subfloat[active]{
    	\includegraphics[width=0.47\linewidth]{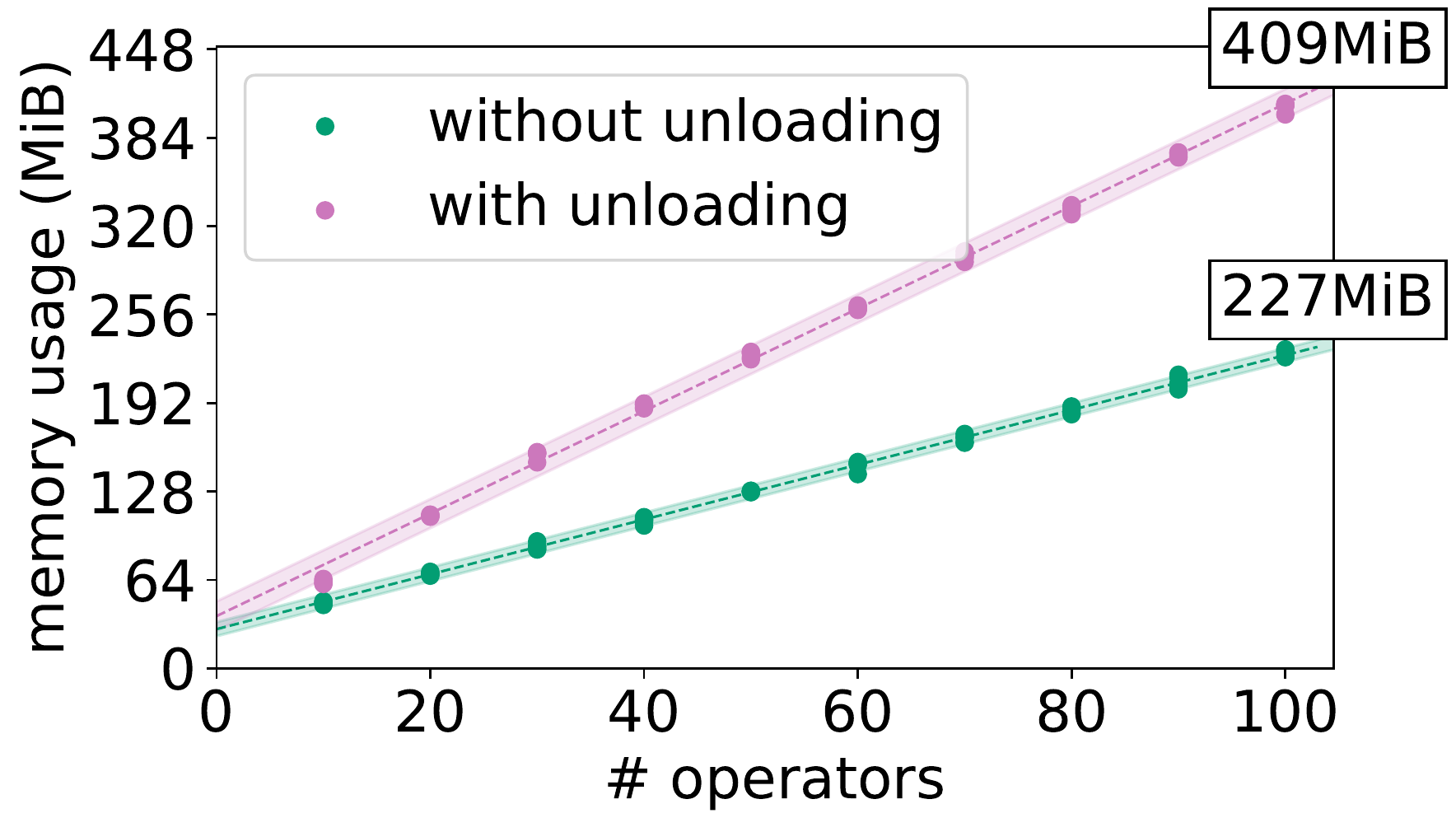}
    	\label{gra:wasm_unloading_active}
    }
    \subfloat[idle]{
    	\includegraphics[width=0.47\linewidth]{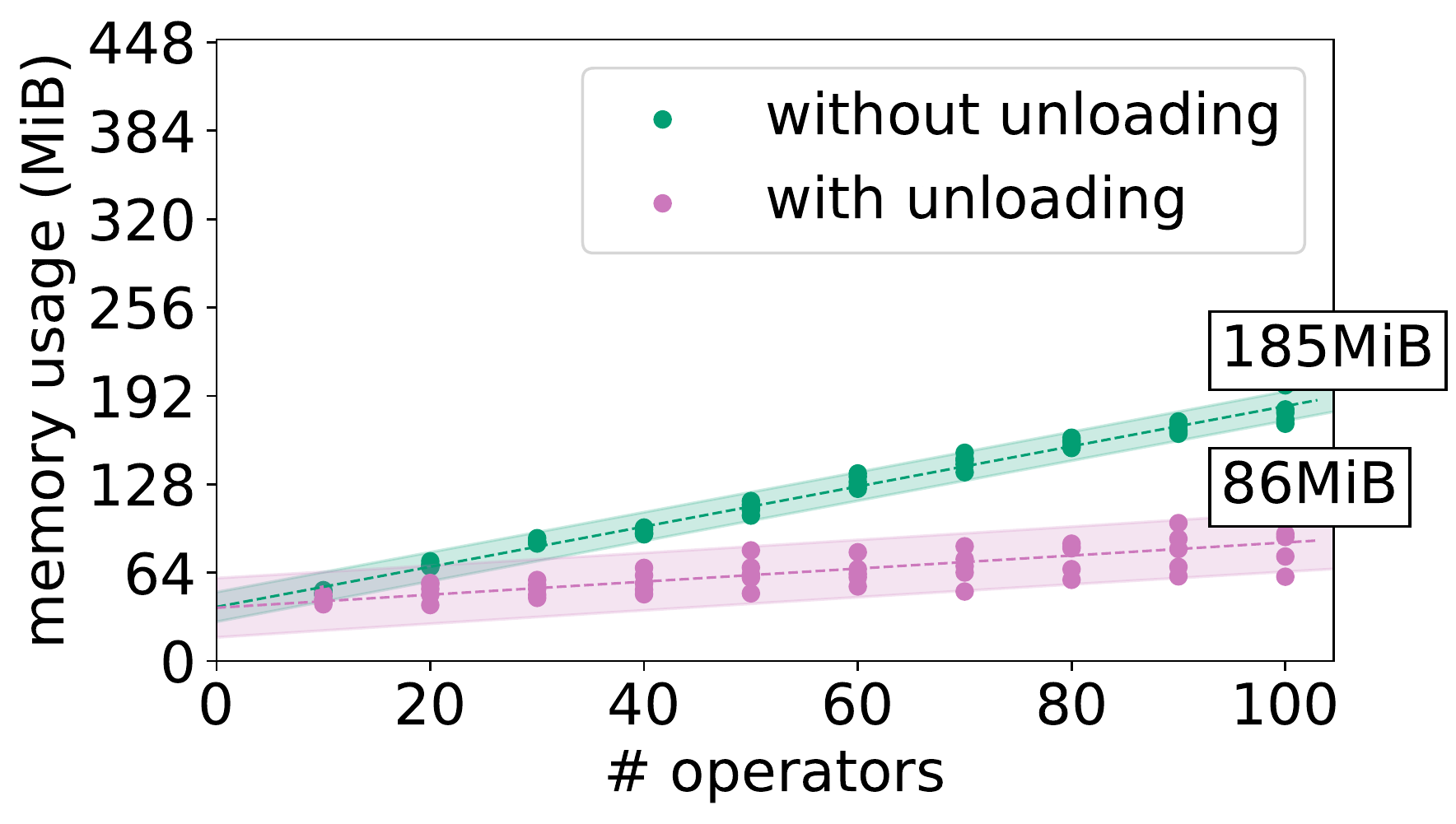}
    	\label{gra:wasm_unloading_idle}
    }
    
    \caption{
        The memory 95\% upper bounds of the WASM operator with automatic unloading enabled/ disabled; very frequent unloading causes more memory usage, for idle operators it can save memory.
    }
    \label{gra:wasm_unloading}
\end{figure}

Figure~\ref{gra:wasm_unloading} shows the obtained memory upper bounds for the synthetic-operator running as WASM modules.
Two versions of the WASM operator are compared: one does not unload the WASM instances and the other unloads each WASM instance in-between each iteration of the reconciliation loop.
The colored areas represent the 95\% prediction intervals for the regression models as described in Section~\ref{result:memory_usage_method}.

\begin{figure}
    \centering
	\includegraphics[width=0.80\linewidth]{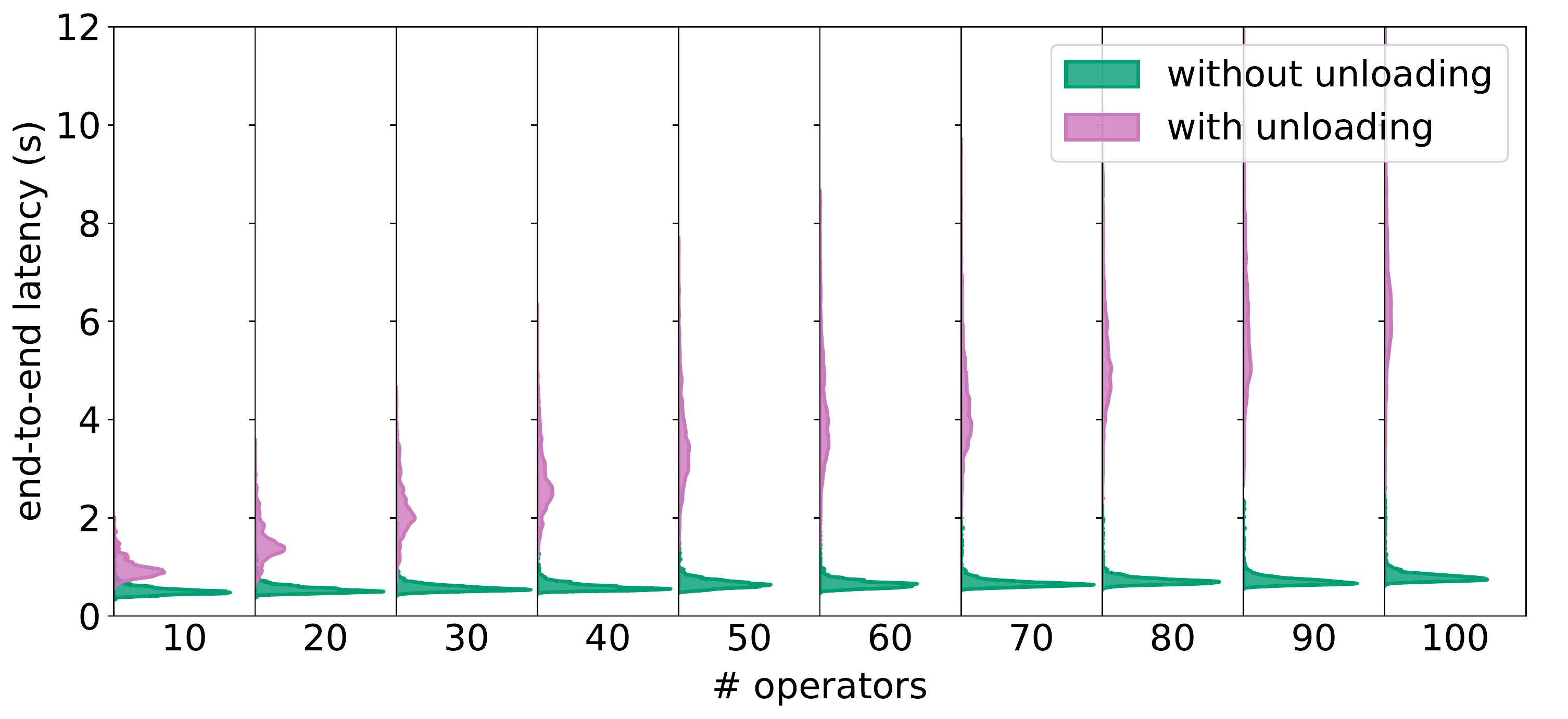}
	\caption{The end-to-end latency for active WASM operator with unloading enabled/ disabled. Actively unloading and swapping modules introduces significant latency.}
	\label{gra:wasm_unloading_e2e}
\end{figure}

Figure~\ref{gra:wasm_unloading_active} shows that the effect of constantly unloading and reloading active WASM operator was an $80.49\%$ increase in memory usage for 100 operators. In Figure~\ref{gra:wasm_unloading_e2e}, the effect of actively unloading and reloading operators on the measured end-to-end latencies is displayed, this figure was obtained as described in \ref{result:e2e_latency_method}. Figure~\ref{gra:wasm_unloading_idle} shows, running 100 operators, we achieved a $52.66\%$ reduction for idle operators compared to not unloading.

Unloading the modules reduces memory usage in case of idle operators. The parent operator writes the memory of idle WASM instances to disk and reloads it later when a Kubernetes watch event is received, as described in Section~\ref{architecture:parent_operator}. Since most operators often stay idle for a long time, this can greatly optimize resource utilization in memory-constrained environments. However, in case of a worst-case unload and reload pattern, memory usage is higher than in case no unloading and reloading takes place. Frequent unloading also introduces a large end-to-end latency penalty due to the disk overhead of swapping the WASM instance, as shown in Figure~\ref{gra:wasm_unloading_e2e}.

To properly benefit from automatic WASM module unloading in a mixed active-idle situation, a predictive scheduler is a necessity, this is considered as future work in this paper. Such a scheduler could help unloading only when it is beneficial to unload a WASM module instead of unloading it in-between each control loop iteration. The optimization opportunity also greatly depends on the heap memory allocated by the operators, necessary for Kubernetes API state caches. This relationship is further discussed in Section~\ref{result:heap_mem}.

\subsection{Dynamically allocated memory}
\label{result:heap_mem}
\begin{figure}
    \centering
    \subfloat[The per-operator memory 95\% upper bounds increase due to an extra 1MiB of dynamically allocated memory for active operators.]{
        \includegraphics[width=0.46\linewidth]{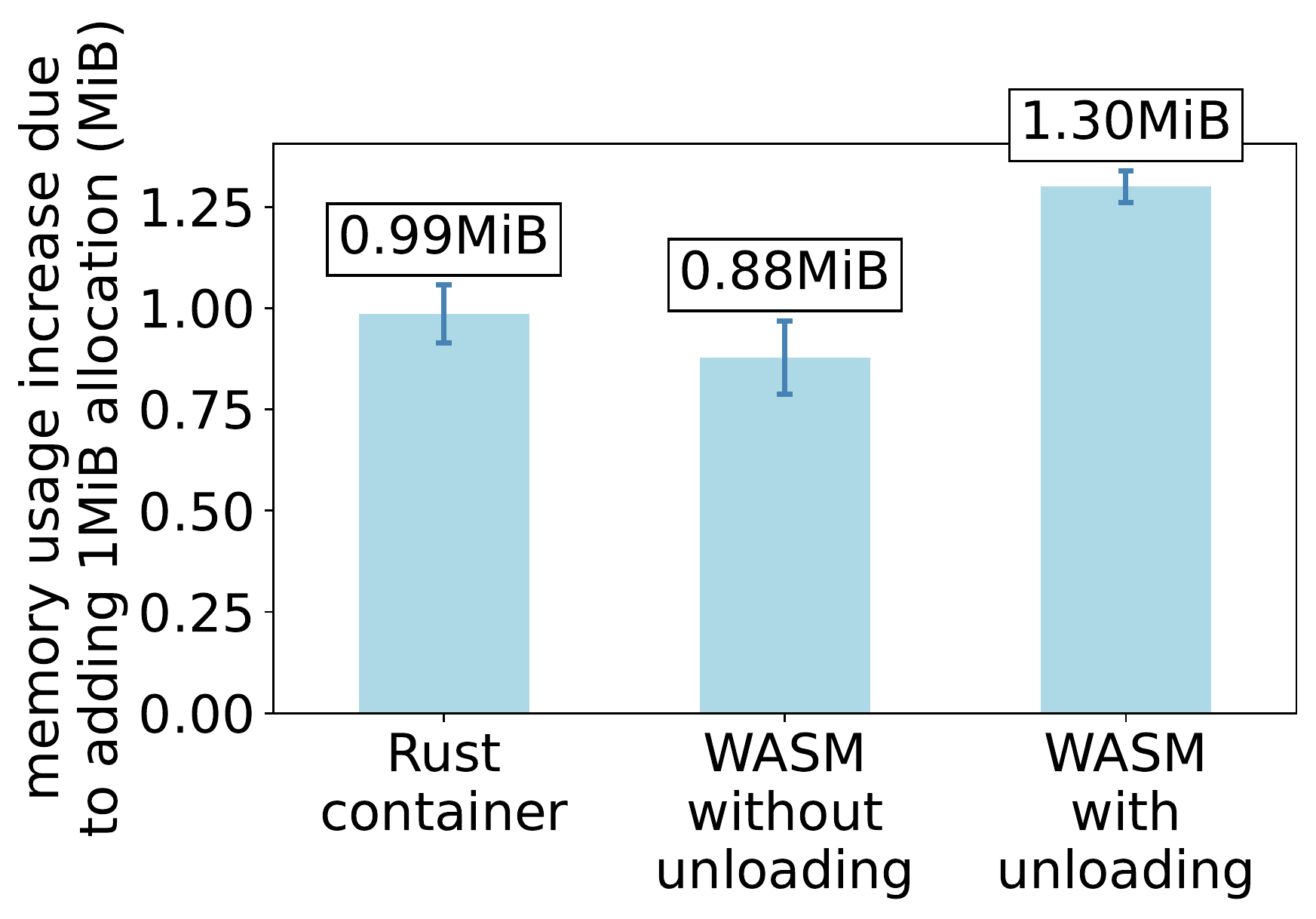}
        \label{gra:heap_mem_active}
    }
    \hspace{1mm}
    \subfloat[The per-operator memory 95\% upper bounds increase due to an extra 1MiB of dynamically allocated memory for idle operators.]{
        \includegraphics[width=0.46\linewidth]{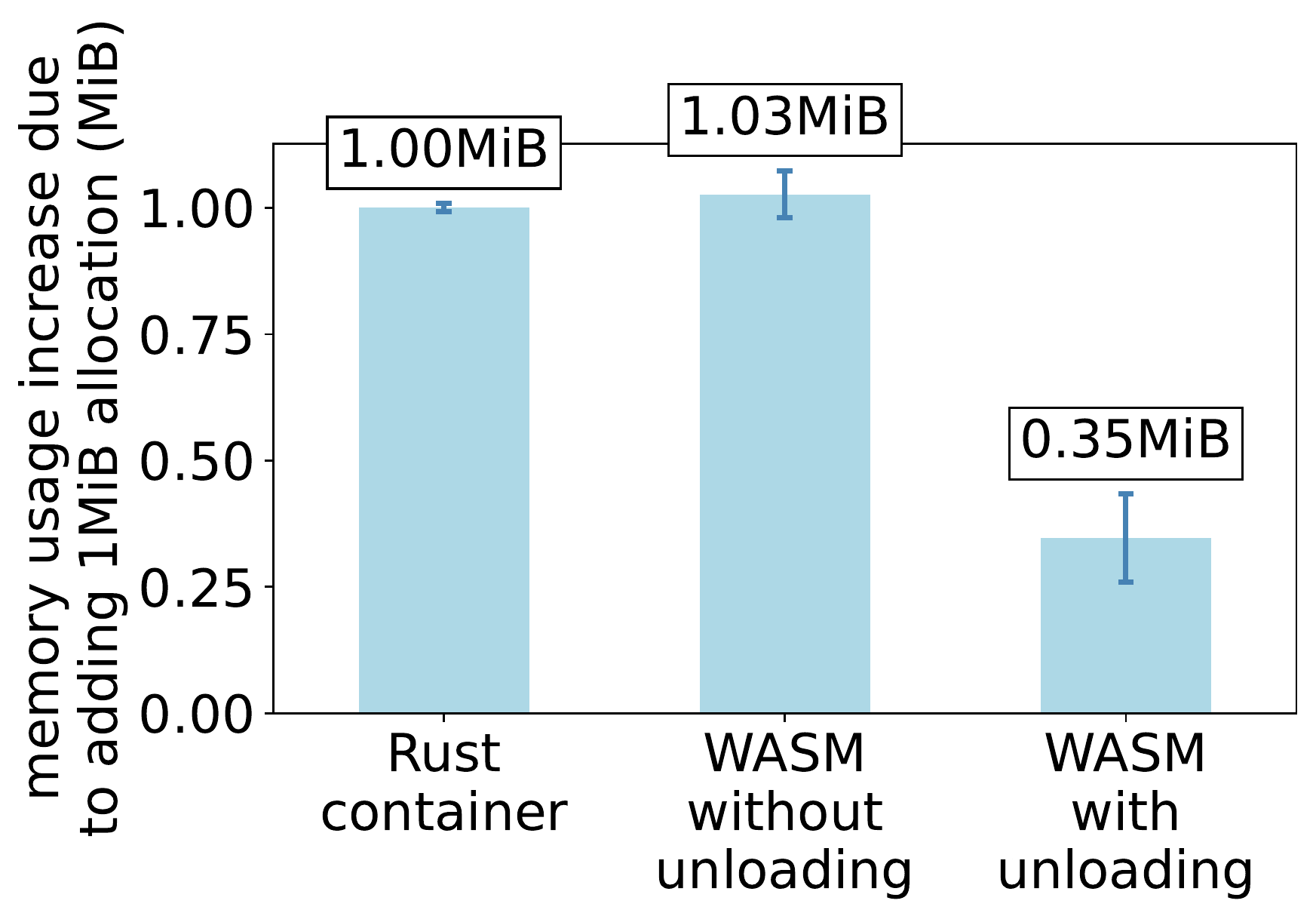}
        \label{gra:heap_mem_idle}
    }

	\caption{The effects of allocating 1MiB of heap memory in each operator on the 95\% upper bounds.}
	\label{gra:heap_mem}
\end{figure}

Figure~\ref{gra:heap_mem} shows the average memory upper bound increase per operator due to a 1MiB increase in dynamically allocated memory. The metric is obtained based on the slope of the linear regression models trained on 20 upper bound memory usage samples obtained for experiments with allocation sizes of 0MiB to 3MiB, with 5 runs per experiments. Also indicated are the 95\% confidence intervals for these slopes.

Figure~\ref{gra:heap_mem_active} shows that dynamically allocating 1MiB additional heap memory in each operator resulted in in a memory upper bound increase of roughly 100MiB for 100 active operators with unloading disabled, and in an increase of 130MiB for active WASM operators with unloading enabled. The 30MiB extra overhead originates from the additional memory required to reload the WASM module.

Figure~\ref{gra:heap_mem_idle} shows that the memory consumption for idle operators only increased with 0.35MiB when using our unloading and swapping solution. This is significantly lower than the memory increases for operators without unloading and swapping.

As discussed in Section~\ref{result:unloading}, adding swapping also adds end-to-end latency. For our experiments, it took about 26ms to swap 1MiB of data to disk per operator, which can be fully attributed to the disk read and write overhead of the hard disk drive in the test server. No latency increase was experienced when using the containerized solution or the WASM solution without unloading.

Operators that watch a large amount of Kubernetes cluster resources will typically keep many of these resources in a cache that they update once the Kubernetes API notifies that a resource change took place. This means that these operators have large amounts of dynamically allocated memory, which directly translates to a memory upper bound increase, as discussed in this section. To reduce this memory usage, it is possible to use our unloading implementation in combination with a tuned scheduler. However, such a solution will result in larger latency overhead due to disk writes. Another solution is to move all operator caches to the parent operator and to deduplicate the resources in these caches.
\section{Conclusion}

Complex Kubernetes operator workloads are often too heavy for constrained environments. In this article, a novel WebAssembly-based Kubernetes operator solution is proposed. This solution demonstrates that WebAssembly, a technology used by edge FaaS solutions, can also be used to reduce the overhead associated with Kubernetes cluster management. It therefore extends the Wasmtime runtime, adding support for asynchronous Kubernetes API interaction and unloading of idle operators. Our test results show a reduction in memory footprint of 100 active synthetic operators from 1405MiB to 227MiB and of 100 idle operators from 1131MiB to 86MiB by using WASM operators instead of traditional operators. This reduction is due to reduced child operator complexity and the lower WebAssembly isolation overhead. We also found that CPU overhead, identified as a drawback of WASM in prior work \cite{Jangda2019}, does not affect end-to-end latency for our synthetic-operator workload. Unloading WASM operators reduces memory usage for idle operators, while increasing memory usage and end-to-end latency for idle operators. Therefore, future work is needed to add a predictive scheduler that fully optimizes this feature.

Our WASM architecture and implementation demonstrate that initiatives, such as the metacontroller project \cite{Yeh2022}, can integrate a WASM runtime as an alternative to their current WebHook solution and benefit from reduced complexity and resource usage. Resource-constrained edge environments are able to run more WebAssembly operators than traditional operators, enabling complex workloads. Cloud deployments become more resource efficient by replacing existing operators with WASM-based operators. The shared benefits of our solution across both edge and cloud segments help accelerate research and adoption.

The biggest open challenges for developing new WASM operators are the WASM and WASI specifications that are still under development. In addition, Golang lacks proper support for WASI, making it more difficult to write operators in Golang. However, Rust operators can more easily take advantage of running as WASM modules. We further propose to obtain additional reductions in memory usage by moving caching logic from the child to the parent operators.

\phantomsection

\bibliography{references}

\end{document}